\documentclass{emulateapj}
\usepackage{psfig}
\usepackage{apjfonts}

\begin{document}

\def\psr{PSR~J1357--6429}
\def\snr{G309.8--2.6}

\slugcomment{Accepted by {\em ApJ Letters}, June 22, 2004}
\shorttitle{THE VERY YOUNG PULSAR J1357--6429}
\shortauthors{CAMILO ET AL.}

\title{The Very Young Radio Pulsar J1357--6429}

\author{F.~Camilo,\altaffilmark{1}
  R.~N.~Manchester,\altaffilmark{2}
  A. G. Lyne,\altaffilmark{3}
  B.~M.~Gaensler,\altaffilmark{4}
  A. Possenti,\altaffilmark{5}
  N. D'Amico,\altaffilmark{5,6}
  I. H. Stairs,\altaffilmark{7}
  A. J. Faulkner,\altaffilmark{3}
  M. Kramer,\altaffilmark{3}
  D. R. Lorimer,\altaffilmark{3}
  M. A. McLaughlin,\altaffilmark{3}
  and G. Hobbs\altaffilmark{2} }
\altaffiltext{1}{Columbia Astrophysics Laboratory, Columbia University,
  550 West 120th Street, New York, NY~10027}
\altaffiltext{2}{Australia Telescope National Facility, CSIRO,
  P.O.~Box~76, Epping, NSW~1710, Australia}
\altaffiltext{3}{University of Manchester, Jodrell Bank Observatory,
  Macclesfield, Cheshire, SK11~9DL, UK}
\altaffiltext{4}{Harvard-Smithsonian Center for Astrophysics, 60 Garden
  Street, Cambridge, MA~02138}
\altaffiltext{5}{INAF --- Osservatorio Astronomico di Cagliari,
Loc. Poggio dei Pini, Strada 54, 09012, Capoterra (CA), Italy}
\altaffiltext{6}{Universit\'a degli Studi di Cagliari, Dipartimento di
Fisica, SP Monserrato-Sestu km 0,7, 90042, Monserrato (CA), Italy}
\altaffiltext{7}{Department of Physics \& Astronomy, University of
British Columbia, 6224 Agricultural Road, Vancouver, B.C. V6T 1Z1, Canada}

\begin{abstract}
We report the discovery of a radio pulsar with a characteristic age of
7300 years, making it one of the 10 apparently youngest Galactic pulsars
known.  PSR~J1357--6429, with a spin period of $P=166$\,ms and spin-down
luminosity of $3.1\times10^{36}$\,ergs\,s$^{-1}$, was detected during the
Parkes multibeam survey of the Galactic plane.  We have measured a large
rotational glitch in this pulsar, with $\Delta P/P = -2.4\times10^{-6}$,
similar in magnitude to those experienced occasionally by the Vela pulsar.
At a nominal distance of only $\sim 2.5$\,kpc, based on the measured
free electron column density of 127\,cm$^{-3}$\,pc and the electron
distribution model of Cordes \& Lazio\nocite{cl02}, this may be, after
the Crab, the nearest very young pulsar known.  The pulsar is located
near the radio supernova remnant candidate G309.8--2.6.
\end{abstract}


\section{Introduction}\label{sec:intro} 

A supernova (SN) explodes in the Galaxy every $\sim 100$ years \citep[see,
e.g.,][and references therein]{cap03,cap04}.  As a result, young
neutron stars in their various guises are rare, as are young pulsars.
Nevertheless, the payoff resulting from their study can be large, in
areas as varied as the physics of core collapse, the internal structure
of neutron stars, and magnetospheric emission processes.  Also, the
youngest neutron stars are often embedded in compact nebulae powered
by relativistic pulsar winds or otherwise interact with their host
supernova remnants --- as such they can make for magnificent probes
of their immediate environment.  Young pulsars are also particularly
useful for establishing reliable birth rates of this important branch
of outcomes of core-collapse SNe.  For these reasons, substantial effort
continues to be devoted to the detection of the youngest neutron stars,
and three-quarters of the Galactic pulsars known with an age $\tau <
10$\,kyr have been discovered in the past seven years at radio and
X-ray wavelengths.

In this Letter we announce the discovery of \psr, a very young,
relatively energetic and nearby pulsar, present its rotational history
during a span of 4.5 years that includes a large glitch, and comment on
the immediate environment of the pulsar and in particular on a nearby
supernova remnant candidate.

\section{Discovery and Observations of \psr }\label{sec:obs}

\psr, with a period $P = 166$\,ms, was discovered on 1999 October 7 in
data collected during the course of the Parkes multibeam survey of the
Galactic plane \citep[e.g.,][]{mlc+01}.  This survey employs a 13-beam
low-noise receiver system at a central radio frequency of 1374\,MHz with
a recorded bandwidth of 288\,MHz.  The large area covered by one pointing
enables the 35\,min-long individual integrations that, together with
the high instantaneous sensitivity, result in the good limiting flux
density of about 0.2\,mJy for long period pulsars over a large area
along the Galactic plane ($|b|<5\arcdeg; 260\arcdeg<l<50\arcdeg$).
In turn this has led to the discovery of more than 700 pulsars
\citep{mlc+01,mhl+02,kbm+03,hfs+04}.

After discovery, as with every newly detected pulsar, we began regular
timing observations of \psr\ at Parkes, using the filterbank-based
observing system employed in the survey.  Typically this consists
of the recording to magnetic tape of raw data for about 15\,min in
order to obtain off-line a pulse profile from which we
derive the time-of-arrival (TOA) of a fiducial point on the profile by
cross-correlation with a high signal-to-noise ratio template.  The average
pulse shape consists of a single approximately symmetric peak with FWHM
= 15\,ms ($0.09 P$).  In this manner we obtained 125 TOAs spanning the
MJD range 51458--53104, a period of 4.5\,yr.

Using the initial set of TOAs together with the {\sc tempo} timing
software\footnote{See http://www.atnf.csiro.au/research/pulsar/tempo .}
we obtained a phase-connected solution for the pulsar accounting for
every turn of the neutron star.  In short order it became apparent that
this pulsar has a large period derivative and therefore a very small
characteristic age, $\tau_c \equiv P/2\dot P = 7300$\,yr.  Some 18 months
after we began timing the pulsar, on about MJD 52000, it underwent a very
large rotational glitch, with a fractional period spin-up of two parts
in one million (see Fig.~\ref{fig:glitch} for the evolution of spin
parameters over time).  In magnitude, this is typical of the glitches
that the $\approx 10$\,kyr-old Vela pulsar experiences every few years
but, interestingly, there is no evidence for an exponentially decaying
component as is observed in the Vela glitches \citep[e.g.,][]{dml00}.

It is not possible to track the rotation phase of the neutron star
across this large glitch.  We find that fitting a timing model to the
full three-year data set since the glitch shows a large quasi-periodicity
in the residuals with a $\sim 500$ day period, probably due to ``timing
noise''.  This can be largely absorbed with a fit requiring a total
of six frequency derivatives to whiten the data.  For simplicity and
to aid observers, the solution we present in Table~\ref{tab:parms} is
based on the most recent one year's worth of data, requiring only two
derivatives, with the second frequency-derivative representing the amount
of timing noise.  We also provide the net jump in spin parameters at the
glitch epoch.  We emphasize that the value of $\ddot \nu$ is {\em not\/}
deterministic (see also Fig.~\ref{fig:glitch}).  It yields an apparent
braking index of rotation $n = \nu \ddot \nu/(\dot \nu)^2 \approx 40$
(where $n$ is defined by $\dot \nu \propto - \nu^n$), which is about 15
times larger than the expected long-term deterministic value.

The presence of significant amounts of timing noise in \psr\ is a
presumed consequence of the less-than-perfect rotational stability of
the young, hot neutron star with a complex superfluid interior
\citep[e.g.,][]{be89,antt94,dm97}.  Such noise biases the pulsar
position obtained through timing methods.  However, we did not have to
rely on timing measurements to obtain the position, since we performed
a pulse-gated interferometric observation with the Australia Telescope
Compact Array (ATCA) on 2000 August 29.  This 5-hour observation used
the array in its 6A configuration, giving a maximum baseline of 6km,
with a central frequency of 1376\,MHz and a bandwidth of 128\,MHz.
The interferometric pulsar position, determined from an image made by
subtraction of an off-pulse data set from an on-pulse data set, is
given in Table~\ref{tab:parms}, and was used for all timing fits.

\section{The Age, Distance, and Vicinity of PSR~J1357--6429}\label{sec:snr}

The actual age of \psr, assuming a constant magnetic moment, is $\tau
= 2 \tau_c [1- (\nu/\nu_0)^{n-1}]/(n-1)$, where $\nu_0$ is the initial
rotation frequency of the neutron star.  We do not know the actual braking
index of the pulsar.  However, if it lies in the range $2 \la n \la 3$,
as is the case for all four pulsars with $n$ measured via phase-coherent
timing \citep[see][and references therein]{ckl+00}, then the real age
is still $\tau \la 15$\,kyr and may be $< 10$\,kyr, depending on the
actual values of $n$ and $\nu_0$.  Also, while such youth is not required
by the occurrence of the large glitch \citep[e.g.,][]{hlj+02}, it is
certainly consistent with it.  There is therefore no doubt that \psr\
is a very young pulsar, one of the $\sim 10$ youngest known in the Galaxy
(see Table~\ref{tab:young}).

\psr\ is also apparently nearby: according to the \citet{cl02} model
for the Galactic free electron distribution and our dispersion measure,
the distance is $d = 2.5$\,kpc.  The \citet{tc93} model yields $d
\sim 4$\,kpc, suggesting the order of the uncertainty inherent in such
estimates.  The implied distance of the pulsar away from the Galactic
plane is 100--200\,pc.  In any case, this pulsar appears to be one of
the two or three nearest to the Earth among those known with $\tau \la
10$\,kyr.  Considering, furthermore, that all known pulsars apparently
younger than \psr\ are associated with a pulsar wind nebula (PWN) or a
supernova remnant (SNR), it seems reasonable to expect that this might
also be the case with \psr.

\begin{figure}
\epsscale{1.00}
\plotone{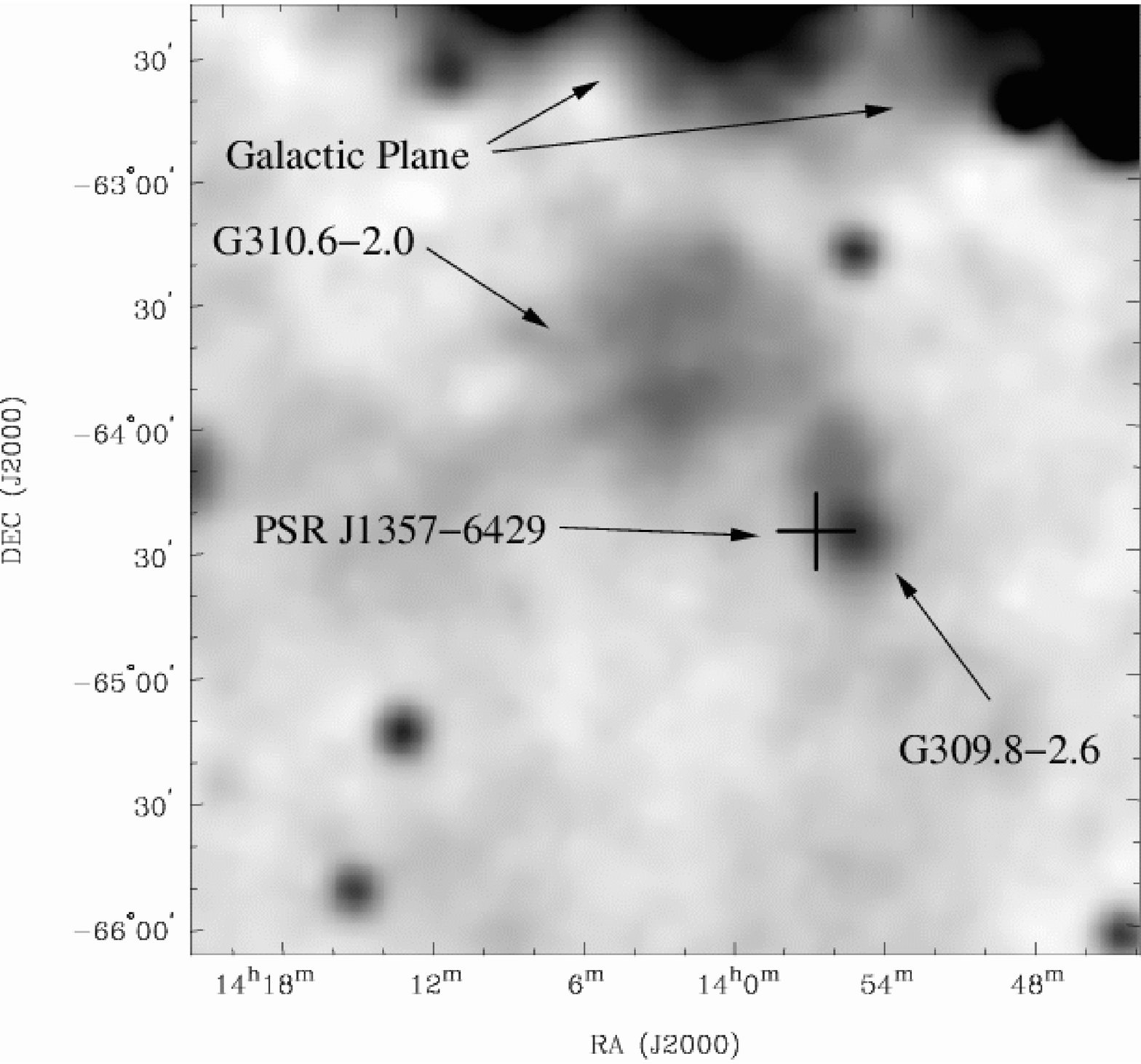}
\caption{ \label{fig:pks_image}
2.4-GHz image of the Galactic plane in the vicinity of \psr, using
data from the Parkes continuum survey of \citet{dshj95}, at a spatial
resolution of $10'$. The greyscale is linear, and ranges between --0.15
and +1.3 Jy~beam$^{-1}$.  The two SNR candidates in this region proposed
by \citet{dshj97} are indicated, while the position of the pulsar
is marked by a cross (the size of the cross is much larger than the
positional uncertainty of the pulsar).  }
\end{figure}

Figure~\ref{fig:pks_image} shows the area around \psr\ as seen in the
2.4\,GHz continuum survey of \citet{dshj95}, with two SNR candidates
proposed by \citet{dshj97} indicated.  It is not clear what, if any, is
the relationship between both SNR candidates, or of either one with a much
larger superimposed shell SNR candidate, G310.5--3.5.  Also, no spectral,
distance or age information is available for these candidates.  The \snr\
SNR candidate appears at this resolution to be about $30'$ in length with a
relatively bright resolved ``core'' to the southwest.  In turn, \psr\
is located, at least in projection, slightly to the east of this core,
but coincident with emission from the SNR candidate.  Apart from \psr,
there are seven other known pulsars located in the area of the Figure,
all old, with $0.5<\tau_c<70$\,Myr.

\section{Discussion}\label{sec:disc}

\psr, with $\tau_c=7300$\,yr, is among the youngest pulsars known in
the Galaxy.  It is also the second youngest pulsar discovered in the
Parkes multibeam survey, after PSR~J1119--6127 with $\tau_c=1600$\,yr
\citep{ckl+00}.  The pulsar is located near the SNR candidate \snr.
While all apparently younger pulsars are associated either with a PWN
or SNR, three slightly older pulsars are not obviously associated with
either such object (Table~\ref{tab:young}).  Further work is therefore
needed in order to investigate the nature of \snr, and also to determine
whether it is in fact associated with the pulsar.

Being a young and energetic pulsar located at apparently only $\sim
3$\,kpc, one would expect substantial X-ray emission to be detectable
from \psr.  As shown in Table~\ref{tab:young}, this is the case for all
younger pulsars (interestingly, only the Crab pulsar among these is a
known optical emitter, and only the Crab and PSR~B1509--58 are known
$\gamma$-ray emitters).  A check of the HEASARC archives reveals no
useful data to address this question, and X-ray observations of this
source will have to be carried out.  Also, there is no known EGRET
$\gamma$-ray source coincident with this relatively energetic and nearby
pulsar \citep{hbb+99}, but \psr\ is likely to be a good target for the
future {\em GLAST\/} mission.

\begin{figure}
\epsscale{0.90}
\plotone{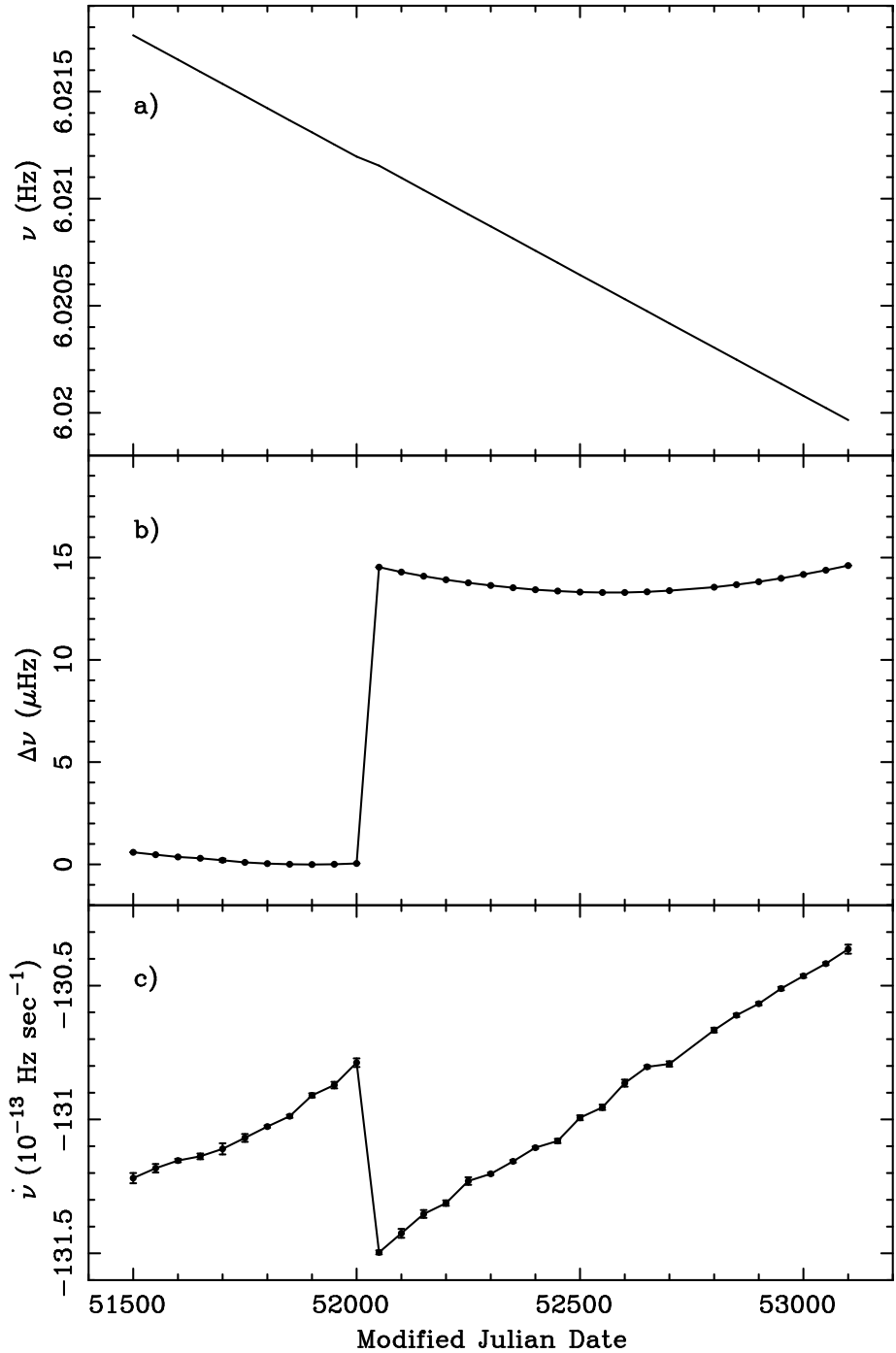}
\caption{ \label{fig:glitch} 
The spin history of \psr.  The frequency ($\nu$) evolution is shown in
panels (a) and (b), the latter with an expanded scale after removal of
$\nu$ and $\dot \nu$ fitted to the 100 days just prior to the glitch that
occurred between MJD~52005 and MJD~52037. The error bars are too small to
see in these panels.  Panel (c) shows the run of $\dot \nu$, obtained from
fits at 50-day intervals using approximately 100-day sections of data.
The discontinuity in $\dot \nu$ at the glitch (0.5\% of $\dot \nu$)
reverses about two years' worth of change in $\dot \nu$ prior to or
after the glitch.  However, these ``continuous'' changes in $\dot \nu$
(the slopes in panel [c]) are not stationary but are rather likely due
to a stochastic process involving the irregular transfer of angular
momentum from the interior to the crust of the neutron star \citep[see
\S~\ref{sec:obs}, \S~\ref{sec:disc}, and also, e.g.,][]{apas84,acp03}. }
\end{figure}

The large value of $\ddot\nu$ (and hence apparent $n$) implied by the
increasing $\dot\nu$ (Fig.~\ref{fig:glitch}) possibly results from
decay after previous (unseen) glitches \citep{jg99,wmp+00}. However,
it is notable that there is no discernible jump in $\ddot\nu$ at the
time of the observed glitch. In most cases, following a large glitch,
a portion of the jump in $\nu$ and $\dot\nu$ decays with a time scale
that ranges from a few days (for the Crab pulsar) to many months (in
the Vela pulsar). This results in a jump in $\ddot\nu$ at the time of
the glitch. The absence of this for the present glitch suggests either
that the decay time-scale is much longer than our data span, or that
different glitches in this pulsar have different characteristics.

The radio luminosity of \psr\ is $L_{1400} \equiv S_{1400} d^2
\sim 3$\,mJy\,kpc$^2$, where $S_{1400}$ is the flux density in the
1400\,MHz band.  This is a rather low value: for the 10 radio pulsars
in Table~\ref{tab:young}, the median $L_{1400} \sim 25$\,mJy\,kpc$^2$.
However, it is in keeping with the realization that the very youngest
pulsars ($\tau \la 10$\,kyr) are not obviously more luminous than
slightly older pulsars \citep[the median luminosity for pulsars with
$\tau_c < 10^5$\,yr is $L_{1400} \sim 30$\,mJy\,kpc$^2$;][]{cam04},
or than middle-aged pulsars \citep[e.g.,][]{cmg+02}.

As we now show, the discovery of a low-luminosity pulsar such as \psr\
contributes greatly toward the estimated birth rate of very young pulsars.
To do this, we make use of the pulsars listed in Table~\ref{tab:young}
that were detected in the Parkes multibeam survey: B1509--58, J1119--6127,
J1357--6429, B1610--50 and J1734--3333. For each pulsar we carried out
a Monte Carlo simulation to correct for the volume of the Galaxy probed
by the multibeam survey, using the $V/V_{\rm max}$-style technique
to estimate the total number of similar objects beaming toward Earth
\citep[for details see, e.g.,][and references therein]{lbdh93}.
Using the most recent estimates of the Galactic pulsar distribution
in these simulations \citep{lor04}, we find that the total number
of such pulsars with $\tau_c < 10$\,kyr is $107\pm95$.  Taking into
account the correction due to the unknown ``beaming fraction'' $f$,
the implied birth rate of this population is $\Sigma = (1.1 \pm
1.0)/f$\,century$^{-1}$ \citep[where $f \approx 0.5$; e.g.,][]{fm93}.
While having a large uncertainty (reflecting the inclusion of \psr,
which accounts for 90\% of the $V/V_{\rm max}$ correction so that it
dominates the calculation; more generally, we may not know enough about
the low end of the radio luminosity distribution), this birth rate is
comparable to those obtained using much larger (and on average much older)
samples \citep[e.g.,][]{lbdh93,lml+98}.

Despite the poorly known pulsar birth rate and beaming fraction,
it appears unlikely that more than $\approx 100$ Galactic pulsars
with $\tau_c < 10$\,kyr are beaming toward us, since the estimated
core-collapse SN rate in our Galaxy (while also having significant
uncertainties) must account also for other branches of neutron star
production \citep[see, e.g.,][for discussions of birth rates of other
types of neutron stars]{bj98,ggv99,gbs00}.  In this case, the 11 pulsars
listed in Table~\ref{tab:young} located on the ``near side'' of the Galaxy
(all but PSR~J1846--0258) could represent about one quarter of all such
young pulsars from which we may ever detect pulsations.  Alternatively,
the discovery of many more would have significant implications for some
of the assumptions underlying these estimates.  Nine of the 12 pulsars
listed in Table~\ref{tab:young} have been discovered in the past seven
years, in both directed and undirected searches, and at radio wavelengths
as well as in X-rays, methods that suffer from significantly different
selection effects.  In view of the above discussion, it is important to
continue with such searches, employing diverse methods.

Young pulsars are rare, and young nearby ones are rarer still.  If \psr\
is indeed located at $d \la 3$\,kpc, with a visual extinction $A_V \la
6$, and with a probable real age of 5--15\,kyr, it is well possible
that its birth event provided a spectacular sight to some of our recent
pre-historical ancestors.  Of greater relevance for us, if not quite
as spectacularly, future study of this very young and nearby pulsar and
any possible PWN/SNR companion may contribute one more significant piece
toward understanding young neutron stars and their environments.

\acknowledgments

The Parkes Observatory and the ATCA are part of the Australia Telescope,
which is funded by the Commonwealth of Australia for operation as a
National Facility managed by CSIRO.  FC acknowledges useful discussions
with Jules Halpern and support from NSF grant AST-02-05853 and a NRAO
travel grant.  IHS holds an NSERC UFA and is supported by a Discovery
Grant.  DRL is a University Research Fellow funded by the Royal Society.


\clearpage

\begin{deluxetable}{ll}
\tablewidth{0pt}
\tablecaption{\label{tab:parms}Measured and derived parameters for
PSR J1357--6429.}
\tablecolumns{2}
\tablehead{
\colhead{Parameter} &
\colhead{Value}
}
\startdata
Right ascension (J2000)\dotfill & $13^{\rm h}57^{\rm m}02\fs43(2)$\tablenotemark{a} \\
Declination (J2000)\dotfill & $-64\arcdeg29'30\farcs2(1)$\tablenotemark{a} \\
Rotation frequency, $\nu$ (s$^{-1}$)\dotfill & 6.0201677726(4) \\
Frequency derivative, $\dot \nu$ (s$^{-2}$)\dotfill 
                            & $-1.305395(4)\times10^{-11}$ \\
Second frequency derivative, $\ddot \nu$ (s$^{-3}$)\dotfill 
                            & $1.16(2)\times10^{-21}$~\tablenotemark{b}      \\
Spin period, $P$ (s)\dotfill        & 0.16610832750(1)      \\
Period derivative, $\dot P$\dotfill & $3.60184(1)\times 10^{-13}$             \\
Epoch (MJD)\dotfill                          & 52921.0             \\
R.M.S. residual (ms) (whitened)\dotfill     & 0.8       \\
Epoch of glitch (MJD)\dotfill                & $52021 \pm 16$ \\
Frequency step at glitch, $\Delta \nu$ (s$^{-1}$)\dotfill
                              & $1.46(1)\times10^{-5}$~\tablenotemark{c}   \\
Change in $\dot \nu$ at glitch, $\Delta \dot \nu$ (s$^{-2}$)\dotfill
                              & $-7.02(3)\times10^{-14}$~\tablenotemark{c} \\
Dispersion measure, DM (cm$^{-3}$\,pc)\dotfill & 127.2(5)           \\
Flux density at 1374\,MHz, $S_{1400}$ (mJy)\dotfill & 0.44(5)              \\
Pulse profile FWHM, $w_{50}$ (ms)\dotfill & 15 \\
Pulse width at 10\% of maximum amplitude, $w_{10}$ (ms)\dotfill & 31 \\
\tableline
Surface magnetic field, $B$ (gauss)\dotfill   
                                                       & $7.8\times 10^{12}$ \\
Characteristic age, $\tau_c$ (kyr)\dotfill       & 7.3      \\
Spin-down luminosity, $\dot E$ (ergs\,s$^{-1}$)\dotfill 
                                                       & $3.1\times10^{36}$  \\
Distance, $d$ (kpc)\dotfill                            & $\sim 2.5$    \\
Radio luminosity at 1400\,MHz, $S_{1400} d^2$ (mJy\,kpc$^2$)\dotfill       & $\sim 2.7 $ \\
\enddata
\tablecomments{Figures in parentheses (except those for position)
are twice the nominal 1\,$\sigma$ {\sc tempo} uncertainties in the
least-significant digits quoted.  The timing fit is based on data
collected over MJD 52737--53104. }
\tablenotetext{a}{Position is obtained from interferometric ATCA
observations. }
\tablenotetext{b}{This parameter is not stationary (see \S~\ref{sec:obs}
for details; see also Fig.~\ref{fig:glitch}). }
\tablenotetext{c}{Assumed glitch epoch for this fit is MJD 52021.  The
error in $\Delta \nu$ is dominated by the uncertainty in this epoch. }
\end{deluxetable}

\begin{deluxetable}{llrccllclll}
\tabletypesize{\scriptsize}
\tablewidth{0pt}
\tablecaption{\label{tab:young}Parameters for all Galactic pulsars known
with apparent age $< 10$\,kyr. }
\tablecolumns{11}
\tablehead{
\colhead{Pulsar}                            &
\colhead{SNR/PWN}                           &
\colhead{$P$}                               &
\colhead{$\log {\rm Age}\tablenotemark{a}$}      &
\colhead{$\log \dot E\tablenotemark{b}$}    &
\colhead{$d$}                               &
\colhead{Wave-}                             &
\colhead{$L_{x\,\rm pwn}\tablenotemark{d}$} &
\colhead{$L_{x\,\rm psr}\tablenotemark{d}$} &
\colhead{$L_{1400}$}                        &
\colhead{Refs.}                             \\ 
\colhead{}                      & 
\colhead{}                      &
\colhead{(ms)}                  &
\colhead{(yr)}                  &
\colhead{(ergs\,s$^{-1}$)}      &
\colhead{(kpc)}                 &
\colhead{band\tablenotemark{c}} &
\colhead{($10^{-3} \dot E$)}    &
\colhead{($10^{-3} \dot E$)}    & 
\colhead{(mJy\,kpc$^2$)}        &
\colhead{}                      \\ } 
\startdata
J1846--0258 & Kes~75      & 323 & 2.9 & 36.9 & 21      & x     & 175  & 8  & \nodata  & (1)\nocite{hcg03} \\
J0205+6449  & 3C~58       &  65 & 2.9 & 37.4 &  3.2    & (x)r  & 0.4  & 0.007 & $\sim 0.5$ & (2)\nocite{shvm04}\\
B0531+21    & Crab        &  33 & 3.0 & 38.7 &  2      & rx    & 29   & 1 & 56  & (3)\nocite{wht+00} \\
B1509--58   & MSH~15--5{\em 2} & 150 & 3.2 & 37.3 & 5  & (x)r   & 8    & 1.7  & 35& (4)\nocite{gak+02}\\
J1119--6127 & G292.2--0.5 & 407 & 3.2 & 36.4 & $\sim6$ & rx    & 0.02 & 0.1 & $\sim 29$ & (5)\nocite{gs03}\\
J1811--1925 & G11.2--0.3  &  64 & 3.2 & 36.8 &  5      & x     & 5    & 0.6 & \nodata & (6)\nocite{rtk+03}\\
J1124--5916 & G292.0+1.8  & 135 & 3.5 & 37.1 & $>6$    & (r)x  & 0.2  & 0.02 & $\ga 3$ & (7)\nocite{hsp+03}\\
J1930+1852  & G54.1+0.3   & 136 & 3.5 & 37.1 & $\sim5$ & (r)x  & 1.7  & 0.3 & $\sim 2$& (8)\nocite{clb+02}\\
{\bf J1357--6429}  & {\bf G309.8--2.6}? & {\bf 166} & {\bf 3.9} & {\bf 36.5} & $\sim$~{\bf 2.5}& r  & ? & ? & $\sim$~{\bf 3} & (9) \\
B1610--50   & ---         & 231 & 3.9 & 36.2 & $\sim 7.8$    & r  &   $<0.7$ & $<0.7$ & $\sim 150$ & (10)\nocite{pkg00}\\
J1617--5055 & ---         &  69 & 3.9 & 37.2 &  $\sim 6.7$    & xr    & ---  & 1  & $\sim 22$   & (11)\nocite{kcm+98}\\
J1734--3333 & ---         &  1169 & 3.9 & 34.7  &  $\sim 7.4$ & r     & ?    & ?  & $\sim 27$   & (12)\nocite{mhl+02}\\
\enddata
\tablecomments{Anomalous X-ray Pulsars and Soft Gamma-ray Repeaters are
not included in this list. For definiteness we also do not include a
few pulsars that may well have $\tau < 10$\,kyr, but for which $\tau_c
\ga 10$\,kyr, such as the Vela pulsar and PSR~J2229+6114 \citep{hcg+01}. }
\tablenotetext{a}{For B0531+21, J0205+6449, and J1811--1925, historical
ages are given assuming associations with SN1054, SN1181, and SN386,
respectively \citep{sg02a}; for other pulsars we list the characteristic
age, $\tau_c \equiv P/2\dot P$. }
\tablenotetext{b}{Spin-down luminosity, $\dot E \equiv 4 \pi^2
I \dot P/P^3$, where $P$ is the pulsar period and $I \equiv
10^{45}$\,g\,cm$^2$. }
\tablenotetext{c}{Pulsar detected in X-rays (x) or radio (r).
Discovery waveband is listed first, and directed searches are denoted
by parentheses. }
\tablenotetext{d}{X-ray luminosities are all given in the 2--10\,keV
band and assume isotropic emission. }
\tablerefs{In each given reference, further references can be found to
the X-ray observations and distance estimates, where appropriate. (1)
\cite{hcg03}; (2) \cite{shvm04}; (3) \cite{wht+00}; (4) \cite{gak+02}; (5)
\cite{gs03}; (6) \cite{rtk+03}; (7) \cite{hsp+03}; (8) \cite{clb+02}; (9)
this work; (10) \cite{pkg00}; (11) \cite{kcm+98}; (12) \cite{mhl+02}. }
\end{deluxetable}

\clearpage

\end{document}